\documentclass[sigconf]{acmart}
\usepackage{booktabs} 
\usepackage[]{graphics}
\usepackage{enumitem}
\usepackage{algorithm}
\usepackage{algpseudocode}
\usepackage{tabularx}
\usepackage{subcaption}
\usepackage[skip=3pt]{caption}

\MakeRobust{\Call}

\algnewcommand{\LineComment}[1]{\State \(\triangleright\) #1}

\renewcommand\footnotetextcopyrightpermission[1]{} 

\begin{document}



\title{Trustworthy Experimentation Under Telemetry Loss}

\newcommand{\affaddr}[1]{#1} 
\newcommand{\affmark}[1][*]{\textsuperscript{#1}}
\newcommand{\affemail}[1]{#1}



\author{Jayant Gupchup}
\affiliation{Microsoft}
\email{jayagup@microsoft.com}

\author{Yasaman Hosseinkashi}
\affiliation{Microsoft}
\email{yahossei@microsoft.com}

\author{Pavel Dmitriev}
\affiliation{Outreach.io}
\email{pavel.dmitriev@outreach.io}

\author{Daniel Schneider}
\affiliation{Microsoft}
\email{danielsc@microsoft.com}

\author{Ross Cutler}
\affiliation{Microsoft}
\email{rcutler@microsoft.com}

\author{Andrei Jefremov}
\affiliation{Unaffiliated}
\authornote{The work was conducted while Pavel Dmitriev and Andrei Jefremov were at Microsoft.}
\email{andrei.d.jefremov@gmail.com}

\author{Martin Ellis}
\affiliation{Microsoft}
\email{maellis@microsoft.com}


%
%


\fancyhead{}

\begin{abstract}
Failure to accurately measure the outcomes of an experiment can lead to bias and incorrect conclusions.
Online controlled experiments (aka AB tests) are increasingly being used to make decisions to improve websites as well as mobile and desktop applications.
We argue that loss of telemetry data (during upload or post-processing) can skew the results of experiments, leading to loss of statistical power and inaccurate or erroneous conclusions.
By systematically investigating the causes of telemetry loss, we argue that it is not practical to entirely eliminate it.
Consequently, experimentation systems need to be robust to its effects.
Furthermore, we note that it is nontrivial to measure the absolute level of telemetry loss in an experimentation system.
In this paper, we take a top-down approach towards solving this problem.
We motivate the impact of loss qualitatively using experiments in real applications deployed at scale, and formalize the problem by presenting a theoretical breakdown of the bias introduced by loss.
Based on this foundation, we present a general framework for quantitatively evaluating the impact of telemetry loss, and present two solutions to measure the absolute levels of loss.
This framework is used by well-known applications at Microsoft, with millions of users and billions of sessions. 
These general principles can be adopted by any application to improve the overall trustworthiness of experimentation and data-driven decision making.
\end{abstract}

   
\maketitle


\section{Introduction}
\label{s:introduction}

AB testing has helped organizations evaluate new ideas, tune parameters, catch critical bugs, predict infrastructure needs, measure customer value, and help with team planning~\cite{fabijan2017benefits, langley2017quic, xu2015infrastructure}.
In the simplest controlled experiment, users are randomly assigned to one of two variants: control (A) or treatment (B). Typically, the control is the existing system, and the treatment is the existing system with a new feature \emph{X}.
If the experiment was designed and executed correctly, the only difference between the two variants is \emph{X}, establishing a causal relationship between the change made to the product and changes in user experience.
The ability to derive this causal relationship is a key reason for widespread use of controlled experiments.
For example, online experimentation helped Bing identify dozens of revenue-related changes to make each month, collectively increasing revenue per search by 10-25\% each year~\cite{ronny2017hbr}.

Although online experimentation is well known in web services, it is also used in mobile and desktop applications~\cite{dmitriev2017tutorial, Xu2016mobileAB}.
Experimentation is especially useful for mobile apps, since they are used by a diverse set of devices and in a wide variety of network conditions.
Such heterogeneity cannot be fully reproduced in a lab, limiting the extent to which applications can be tested internally.
We refer to such experimentation scenarios as \emph{client experiments}, to distinguish them from experiments that purely impact server-side behavior.

One important aspect of client experiments is that the telemetry data collection is done over the Internet, with clients uploading experiment data to cloud services.
This introduces several possible sources of failure, with telemetry data potentially being delayed, lost, or collected at different rates among variants (e.g., due to bandwidth limitations or software bugs).
Moreover, since incoming telemetry data are typically processed in complex data pipelines, there are further opportunities for bugs to be introduced, resulting in missing experiment data; note that all online experiments are susceptible to this risk, not just client experiments.
The key observation is that when telemetry loss is not uniform at random, experiments can be exposed to major population bias, increasing the risk of incorrect conclusions.

To help illustrate the direct impact of telemetry loss on experimentation, we will use an example from one of our recent experiments.
This experiment (which we will refer to as the \emph{ui-change} experiment) involved evaluating the impact of a change to the user interface used to 
survey user satisfaction in Skype, one of our communication products.
After collecting data from the experiment, we found a surprisingly large improvement of 8\% in user ratings.
However, further investigation revealed a software bug, where clients in the treatment group failed to submit poor ratings, resulting in a 13\% difference in telemetry loss compared to the control group.
After fixing the issue and repeating the experiment, we found no statistical difference between the interfaces.

In the previous example, telemetry loss was fairly easy to detect; however, there are also more subtle cases, where the impact may not be so obvious.
For example, in cases where the treatment has different effects across the overall population, this can combine with non-uniform telemetry loss across the population to obscure the results of the experiment.
Specifically, telemetry loss can result in under-representation of substantial segments of the population where the treatment has effects, leading an experimenter to incorrectly conclude that there is no overall difference.
An analogy to this situation can be found in political polling, where polling firms might under-sample certain segments of the voting population; by surveying a non-representative subset of the electorate, they miss important shifts in voting intention and therefore 
make incorrect predictions about election results~\cite{keeter2006impact, celli2016predicting}.


Our experiences clearly demonstrate the importance of reducing telemetry loss; however, eliminating this loss completely (although desirable) is not a realistic goal for large-scale client applications.
This demands a robust methodology to ensure the trustworthiness of experimentation and data-driven decisions made in the presence of such losses.
In this paper, we describe the methodology and tools developed at Microsoft to achieve this goal.
Considering that the impact of such loss is amplified in client experiments, we use specific in-depth examples from two widely used applications; \emph{Skype}, a popular communication application, and \emph{OneNote}, a widely used application for free-form note taking and collaboration.
Note, however, that the proposed solutions are generic and can be applied to any online experimentation system (client- or server-based).
To the best of our knowledge, our work is the first to address telemetry loss in online experimentation.
The main contributions of this paper are as follows:
\begin{itemize}[leftmargin=*,noitemsep,topsep=1pt]
  \setlength\itemsep{0.2em} 
  \item A taxonomy of telemetry loss scenarios and best practices to minimize the loss. 
  \item Examples of real experiments conducted at our organization illustrating the biases introduced by data loss.
  \item A theoretical breakdown of biases caused by data loss and their properties in practice.
  \item A methodology to simulate experiment results under no loss. 
  \item A framework to estimate how much data loss can be tolerated by the experimentation system for trustworthy operation.
  \item Two methods for measuring data loss and clear guidance on their application.
\end{itemize}

The rest of the paper is organized as follows:
Section \ref{s:background} provides detailed background and related work.
Section \ref{s:exp-under-loss} outlines a systematic framework for characterizing loss and estimating the level that can be tolerated by an experimentation system.
Section \ref{s:dataloss} describes and evaluates two novel approaches for measuring telemetry loss, and recommends ways to reduce it.
Section \ref{s:guidance} shares some practical lessons for the benefit of the experimentation community, and Section \ref{s:conclude} summarizes the paper and discusses future work.

\section{Background}
\label{s:background}
\subsection{Related Work}
\label{s:related}
Online controlled experimentation is an active research area, focused on topics such as running controlled experiments in practical settings~\cite{kohavi2013rules, kohavi2009guide, dmitriev2017metrics}, and new statistical methods to improve metric sensitivity~\cite{deng2013sensitivity}.
For an in-depth introduction to online controlled experiments, see~\cite{tang2010overlapping, dmitriev2017tutorial}.
These studies provide context for our work, sharing algorithms, methods, and lessons learned from running real experiments in practice. 
Specifically, telemetry loss was listed as a common reason for misinterpreting the results of experiments~\cite{dmitriev2017metrics}. 

A common technique to measure packet loss in the Internet is through the use of \emph{sequence numbers} (e.g., in the RTP protocol~\cite{schulzrinne2003rfc}).
The sequence number is incremented for each data packet sent, allowing the receiver to detect packets that do not arrive.
We apply a similar approach (referred to as the sequence method) to measure data loss, which we describe in Section \ref{subs:sequence_method}.
The challenges in mirroring data between unreliable endpoints have been studied in detail by the database community~\cite{gray1996dangers, wiesmann2000database}. 
While telemetry gathered from our apps do not require transaction guarantees, mobile environments have constraints in terms of network reliability and local storage.
Moreover, most analysis systems do not tolerate delays of more than a day. 
Nevertheless, client apps can borrow ideas such as lazy replication to improve telemetry reliability, especially if the experimentation and analysis system can tolerate longer delays. 

Although telemetry data loss is recognized as an important problem, there is little prior research discussing data loss in the context of online experiments.
A similar problem of missing treatment values in behavioral randomized experiments is discussed in~\cite{Molinari2010MissingTrt}.
This method assumes no data loss for outcome variables but possible loss of received treatment indicators. 
Our study focuses on the situation where outcome variables and possibly covariates are lost but the treatment assignment is known. 
%
A methodology to identify treatment effects under data loss for categorical outcome metrics in the context of voting experiments is described in~\cite{Imai2009}.
This study is relevant to our work, and captures one of the biases characterized in this paper; however, we formalize the different types of biases induced by data loss, and provide a general framework to establish a tolerance threshold for AB test systems operating at scale.
  
\setlength{\belowcaptionskip}{0pt}
\begin{figure}[t]
  \centering
  \includegraphics[width=\linewidth]{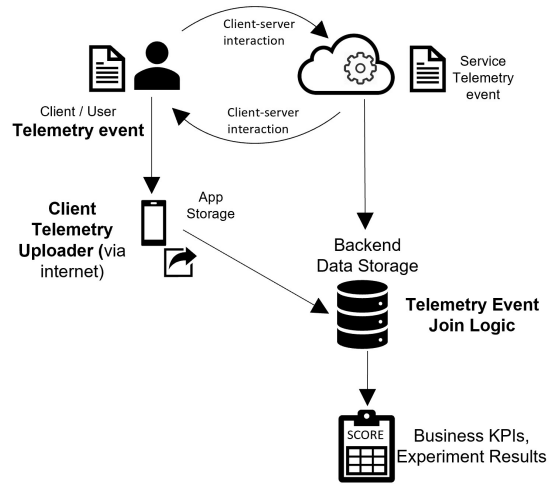}
  \caption{Telemetry flow for the client and server components in a typical application.}
\label{fig:telemetry_flow}
\end{figure}


\subsection{Overview of client telemetry loss}
\label{subs:telemetry_loss}
The processing of client telemetry data collected from apps is a complex and tedious process.
In order to highlight these complexities, we first present a high-level overview of a typical telemetry flow.
Figure \ref{fig:telemetry_flow} shows the interaction between the client and server components and the telemetry flow within an application.
We refer to each client-server interaction as a session.
While we use \emph{Skype} to highlight the details of the complexities, many of the issues are general and apply to \emph{OneNote} and other client applications.

In \emph{Skype}, telemetry data (referred to as events) for a call (or session) are collected at each participating client and at the server. 
These events are sent to the backend store using an uploader application.
The server and backend storage are typically co-located in the cloud so the network connection is reliable and buffering capacity is not a constraint.
The client events are cached locally and transmitted to the backend store opportunistically by the uploader.
Once events arrive at the backend, they are combined to form a call record (or session record) table, which is used for tracking business metrics and experimentation analysis.
Next, we will look at each component and the sources of telemetry loss in more detail.

\noindent
\textbf{Telemetry Events:} In an application, events are reported by a number of internal components interacting with one another. 
For instance, in \emph{Skype}, the call signaling component and service telemetry provide information about call establishment, the audio/video (AV) components provide telemetry about AV quality, and finally, the UI layer reports the overall quality of experience as reported by the user.
This user rating event (previously discussed in reference to the \emph{ui-change} experiment) is not collected for every session, but is available only for a subset of randomly selected calls, when users are asked to rate their experience. 
In contrast, call setup telemetry (henceforth, \emph{CST}) is collected for every call attempt.
The \emph{CST} event is used to compute key metrics such as call establishment rate and call duration.
It is worth mentioning that every event also carries experiment configuration information (i.e., treatment assignment identification).
This information is used to generate \emph{experiment scorecards}, lists of metric comparisons to help track the impact of treatments on user experience, as well as metrics for experiment and data health.
Typically, each component generates one event per session, which are cached locally and uploaded to the backend using an uploader application. 
Henceforth, the details within an event will be referred to as ``measures''. 

\noindent
\textbf{Storage and Event Uploader:} 
Challenges faced by the telemetry uploader application, which can result in data loss, include:

\begin{itemize}[leftmargin=*,noitemsep,topsep=1pt]
\setlength\itemsep{0.2em} 
\item{Bandwidth heterogeneity: Clients make calls (or initiate sessions) in a range of network conditions, including metered 2G networks. 
These bandwidth constraints result in poor event transmission reliability.} 
\item{Sharing bandwidth with the service: Sending telemetry during a session can impact the session experience, so the telemetry system needs to be service-aware, and back off to prioritize smooth functioning of the app.}
\item{App termination: Once an app has been moved to the background, it can be abruptly terminated by the operating system or the user. This prevents some state from being persisted to disk.}
\item{Limited storage budget: Typically, the local cache is implemented as a buffer with a limited storage budget (e.g., on low-end mobile devices). 
If the telemetry queue exceeds the allocated storage, events are dropped.}
\item{Event prioritization: Events are prioritized to transmit the most critical information first. Business KPIs (e.g., user ratings) are sent first, which may cause starvation of lower priority events 
(e.g., technical metrics).
Typically, only the highest priority events are sent reliably (with re-transmission enabled) while the lower ones are not.
It is worth noting that many component-level experiments rely on metrics carried by lower priority events.}
\end{itemize}

As shown in Figure \ref{fig:telemetry_flow}, events are collected from both client and server, which upload events independently.
This can result in telemetry events from the same session arriving at different times, and suffering different rates of data loss.

\noindent
\textbf{Event Joins:} Once events are uploaded to the backend system, they are joined together to form the call record (or session record) that powers experimentation analysis.
Typically, the join key is $\{session\_id, endpoint\_id\}$.
Since the events are uploaded separately, we may receive some events, while others may fail or suffer delays.
If a component fails to record $endpoint\_id$ or $session\_id$ correctly due to software bugs, those events will be dropped downstream due to the join failure.
In the \emph{Skype} scenario, each call has a unique ID and the join key is $\{call\_id, endpoint\_id\}$


\subsection{Biases caused by telemetry loss} 
\label{subs:loss_experimentation_examples}
In this section, we describe real experiments in \emph{Skype} and \emph{OneNote}, illustrating different 
types of bias caused by telemetry loss. These experiments were run at scale having more than a million
observations in each variant.

\textbf{\emph{OneNote} First-Run Experience - Differing loss rates within segment: }
This experiment tested the number of swipe screens after the installation of the app (referred to as the ``first-run experience''). 
The control experience showed the user three swipe screens, while the treatment only showed one.
The hypothesis was that three swipe screens was too many for some users, and that reducing it to one would yield higher engagement with the app.
The AB scorecard for the first-run population showed a statistically significant difference in the number of users in treatment and control,
a phenomenon known as sample ratio mismatch (SRM) ~\cite{dmitriev2017metrics}. The SRM was found to be around $4\%$, with more users in the treatment compared to the control.
Furthermore, within the population of users running the app for the first time, the scorecard suggested that the users in the treatment group had lower engagement.
The overall scorecard computed on the entire population, on the other hand, did not show an SRM, showing instead that users in the treatment group had higher engagement with the app.
The SRM in the first-run scorecard was caused due to a bug in the control group, resulting in a failure to submit telemetry in cases where users abandoned the first-run screen.
In this case, telemetry loss led to a biased estimate of the engagement metric in the scorecard.
After the bug was detected and fixed, the experiment was repeated, showing no improvement in either the overall population or the first-run population.

\textbf{\emph{Skype} Video Bandwidth - Differing loss rates among variants: } 
This experiment aimed to improve video quality in low bandwidth conditions.
It had four treatments, each testing a different setting. 
We found that treatment groups with a lower bandwidth threshold were exposed to higher data loss rates due to more challenging conditions.
The data loss rate in each setting was found to be statistically different; the difference in data loss between the variants with the highest and lowest loss rates was $0.7\%$.
In this experiment, the data loss of the event carrying the outcome metric was directly correlated with the treatment itself, (i.e., the treatment has an impact on the data loss rate).
The unbalanced loss rates between treatment and control variants resulted in a biased comparison, ultimately invalidating the results of the experiment.

\textbf{\emph{Skype} Video Decoder - High absolute loss rates across variants: }
This experiment 
focused on improving video quality for mobile clients. 
Initial analysis did not show any significant difference between control and treatment, and there were no apparent problems with data quality.
However, further investigation showed that the video telemetry had a loss rate over $10\%$, since the video telemetry event had a lower priority (as discussed in Section \ref{subs:telemetry_loss}), and suffered the highest loss rates compared to other client events.
Since none of the outcome metrics had changed between treatment and control, it was unclear whether ``no change'' was due to telemetry loss or truly due to the treatment having no impact. 
From this experiment, we learned that, even when the loss rate between control and treatment is balanced, a very high overall loss rate can lead to massive bias in outcome metrics, masking the true impact of the treatment on user experience. 
In such cases, the scorecards show an incomplete picture about the effect of the treatment, and can lead to inaccurate conclusions.
This topic will be discussed in more detail in Section \ref{s:exp-under-loss}.

\textbf{\emph{Skype} Headset Impact - Unable to construct segments: }
In this experiment, we were interested in improving the audio experience of users using headsets.
The outcome metric was the average user rating, which comes from a reliable event with low loss rate.
However, for each variant, we also needed to identify whether a headset was used during the call, which relies on device usage information reported from a low priority event with high loss rate. 
As a consequence, analysis of this experiment was biased due to the loss of the information needed to identify the sub-populations of interest.

The examples and scenarios presented above are representative of client experiments.
As a result, it is critical to consider telemetry loss of events at the design phase of the experiment.
Our approach was to look at this problem in a top-down fashion.
We first begin by evaluating the impact of loss on experimentation to understand how much loss can be tolerated.
Then, we address the challenge of measuring the absolute level of loss for each event to assess how far we are from our desired loss target. 
As mentioned before, our solutions are general and apply to any experimentation system; we use the context of \emph{Skype} to convey these ideas more concretely since this app is used in challenging environments.

\section{Experimentation under Data Loss}
\label{s:exp-under-loss}
The presence of telemetry loss in experimentation data raises some natural questions. In particular,
\noindent
\begin{enumerate}[leftmargin=*,noitemsep,topsep=1pt]
  \item How different would the current experiment results be if there was no telemetry loss?
  \item How much telemetry loss can our experimentation system tolerate, while still providing trustworthy results?
\end{enumerate}
Addressing the first question is crucial to ensure the correctness of decisions made by experiments impacted by telemetry loss.
By answering the second question, we can determine a practical goal for improving telemetry loss.
To quantify the impact of telemetry loss, we develop a model to formulate the biases caused by loss on AB test results and explore its practical implications.
Then, we present a practical algorithm to simulate the results of an experiment under no telemetry loss.
Finally, present a general algorithm to estimate a threshold for telemetry loss that can be tolerated in a trustworthy experimentation system.
This algorithm can be adapted and tuned for different experimentation platforms with variable sensitivity to telemetry loss.

\subsection{Statistical Model for Data Loss Impact} 
\label{subs:model}
The change in user experience caused by the treatment is commonly referred to as ``treatment effect'' in AB testing.
Without loss of generality, we describe the mathematical models in this section for a simple AB test with two variants: control and treatment.
Following the linear model framework for randomized experiment ~\cite{montgomery2008design}, if no blocking variable is present, then treatment effect ($\beta_T$) is modeled as the difference in expected outcome variable $Y$ conditioned on treatment $T$ (an indicator variable representing treatment):
\begin{eqnarray}\label{model1}
Y &=& {\beta}_0 + {\beta}_TT + \varepsilon	\\
\beta_T &=& E(Y|T=1) - E(Y|T=0) 
\end{eqnarray}
assuming $\varepsilon$ has mean $0$ and finite variance $\sigma^2$.

In this framework, $\beta_0 = E(Y|T=0)$ represents the baseline or current average of the outcome metric while $\beta_T$ measures its expected change if treatment is applied to all population.
Under fairly general assumptions, the difference between the averages of $Y$ in treatment and control,
\begin{eqnarray}\label{delta}
\Delta = \bar{Y}_1 - \bar{Y}_0
\end{eqnarray}
\noindent
is an unbiased estimator of $\beta_T$ and its normalized value $\frac{\Delta}{se(\Delta)}$ offers asymptotic unbiased significance tests.
This is widely used in online and client experimentation as a reliable decision making mechanism.
However, this solution is no longer adequate in the presence of telemetry loss, where model (\ref{model1}) needs to be extended by a new term to capture the loss.
Let $L\sim Bin(p_{_{T,X}})$ be $1$ whenever outcome variable $Y$ is lost and $0$ otherwise.
Then model (\ref{model1}) under data loss extends to:
\begin{equation}\label{modelLoss}
Y = {\beta}_0 + {\beta}_TT + {\beta}_LL+ \beta_{int} T\times L + \varepsilon
\end{equation}

According to (\ref{modelLoss}), $E(Y|_{T=1}) = \beta_{0} + \beta_T + \beta_L E(L|_{T=1}) + \beta_{int}E(L|_{T=1})$ 
and $E(Y|_{T=0}) = \beta_{0} + \beta_L E(L|_{T=0}) )$.
Therefore,
$$E(\Delta) = \beta_T + bias(\Delta); where $$
\begin{equation}\label{biasFormula1}
bias(\Delta) = \beta_L \times [E(L|_{T=1}) - E(L|_{T=0})] +\beta_{int} E(L|_{T=1})
\end{equation}

The two additive terms in (\ref{biasFormula1}) are the two types of biases with very different characteristics in practice: \emph{correlation bias} and \emph{interaction bias}.

\textbf{Correlation Bias (corr-bias):} $\beta_L \times (E(L|T=1) - E(L|T=0))$, measures the correlation between data loss and treatment.
This is non-zero when the treatment changes the behavior of the app in sending telemetry, or causes an indirect change in the loss distribution.
For example, the \emph{Video Bandwidth} experiment described in Section \ref{subs:loss_experimentation_examples} exhibits corr-bias.

\textbf{Interaction Bias (int-bias):} $\beta_{int} E(L|T=1)$, measures the interaction between treatment effect and data loss.
This occurs when the treatment effect is expected to be different under $L=1$ and $L=0$.
For example, the \emph{Video Decoder} experiment described in Section \ref{subs:loss_experimentation_examples} exhibits int-bias.

Under the missing at random (MAR) assumption, $L$ is independent of $T$ and $Y$; i.e., $E(L|_{T=1}) = E(L|_{T=0})$ and $\beta_{int} = 0$.
Hence, both bias terms would be zero and observed delta in Formula (\ref{delta}) still provides an unbiased estimator of true impact.
The only cost of telemetry loss in this case is reduced sample power.
In practice, however, we observed that telemetry loss, even when occurring at the same rate for treatment and control group, is not independent of the outcome variable or treatment itself.

In the \emph{Video Bandwidth} experiment, more challenging bandwidth conditions led to a higher loss rate.
This resulted in a positive correlation between data loss and treatment, for video metrics.
Since the data loss is not at random and is more concentrated on poor user experiences (i.e., low bandwidth is closely related to poor user experience~\cite{jiang2016via}), there was no way to separate the impact of treatment from the inherent differences in samples.
Therefore, the video metrics reported in our scorecards were inconclusive.

Correlation bias manifests itself as statistically different loss rates between treatment and control samples, so it is fairly easy to detect.
However, there is no easy way to recover from it, since changes in the outcome metrics could be attributed to either the treatment change or the telemetry loss.
If the correlation bias is a result of a bug, such as in \emph{ui-change} example, the solution is to fix the bug and run the experiment again.
However, if the correlation bias is because of the nature of treatment, such as in \emph{Video Bandwidth} experiment, the only option is to analyze the experiment using other metrics that are not impacted.

In contrast, interaction bias is difficult to detect and correct for.
This hidden bias can easily mislead experimenters who observe the same rate of telemetry loss on both treatment and control side and conclude there is no obstacle to analyzing the experiment data.

\subsection{Challenges in Correcting Data Loss Bias}
\label{subs:post-strat}
Covariate post-stratification is a common statistical technique to adjust estimates of an outcome variable by re-weighting~\cite{little2014statistical}.
This method is especially beneficial in handling data loss if there is a random variable $X$ that is strongly correlated with loss, and the true distribution of $X$ is observed.
This adjustment can be achieved by dividing the population into strata with known weights, imputting the lost values using the correlated feature, and finally re-computing the outcome metric using the strata weights.

In practice, applying this method is problematic and finding a reasonable covariate is non-trivial.
In \emph{Skype}, when data loss occurs due to extreme network conditions, obtaining accurate network estimates is challenging, and therefore correlated covariates are not easily available.
Moreover, data loss is a nonlinear function of multiple variables, as discussed in Section \ref{subs:telemetry_loss}.
Another limitation of this method is faced when one or more strata are completely lost, such as application crashes which lead to $100\%$ data loss.
In such cases, knowledge of loss rates does not help to correct for the bias.
%
In the absence of suitable covariates, we have developed solutions to estimate boundaries of data loss bias instead of correcting for it.

\setlength{\textfloatsep}{10pt}
\begin{algorithm}[t]
\caption{Simulate Overall Treatment Effect}
\begin{algorithmic}[1]
\Require $[\bar{y}_{ctrl}'', s_{ctrl}'', \beta_{int}], [\bar{y}'_{ctrl}, s'_{ctrl}, s'_{trt}, \bar{y}'_{trt}], [l_{ctrl}, l_{trt}]$
\Ensure overall treatment effect under no loss: $\frac{\Delta}{se(\Delta)}$
\State  Estimate sample mean for lost data points according to  (\ref{e:unknown-mean})
\State  Estimate overall delta:  
	\begin{equation}\nonumber
      \Delta = (1-l_{ctrl}, 1-l_{trt})\times \left(
                                              \begin{array}{c}
                                                \bar{y}^\prime_{ctrl} \\
                                                \bar{y}^\prime_{trt} \\
                                              \end{array}
                                            \right)
      +(l_A,l_B ) \times \left(
                                              \begin{array}{c}
                                                \bar{y}^{\prime\prime}_{ctrl} \\
                                                \bar{y}^{\prime\prime}_{trt} \\
                                              \end{array}
                                            \right)
    \end{equation}
\State Calculate overall variances $s_{trt}^2$ and $s_{ctrl}^2$ according to (\ref{e:sbreakdown2}) 
\State Calculate overall treatment effect $\frac{\Delta}{se(\Delta)}$  where $se(\Delta)$ represents the  standard deviation of delta
\end{algorithmic}
\label{a:simulation1}
\end{algorithm}

\subsection{Estimating Boundaries of Data Loss Bias}\label{subs:simulation}
\label{s:boundaries}
In this section, we address how the results of experiments would be different if telemetry was not lost.
Simulation is a powerful tool for this, since it does not require external data or covariates correlated with loss $L$.
Beyond observed summary statistics, we need only the measured loss rates for each sample and a range of \emph{scenarios}. 

A \emph{scenario} characterizes an intuition about the lost data points, such as ``lost events are correlated with poor user experience, but the experience is improved by the treatment'' or ``lost events are correlated with poor user experience, and the treatment degrades the experience even further''. 
The idea is to impute the lost data points under certain scenarios and then reconstruct the overall treatment effect.
This can be done by decomposing the mean and standard deviation of a complete sample $(\bar{y}, s)$ into observed and unobserved pieces.
We use $(\bar{y}^{\prime}, s^\prime)$ and $(\bar{y}^{\prime\prime}, s^{\prime\prime})$ to refer to the summary statistics of observed and lost parts, respectively.
Here is the breakdown for overall sample mean $\bar{y}$ and variance $s^2$:
\begin{eqnarray}
\bar{y} &=& (1-l)\bar{y}^\prime + l\bar{y}^{\prime\prime}, and\label{e:meanbreakdown}\\
s^2 &=& \frac1n \sum_i (y_i - \bar{y})^2\nonumber\\
&=&\frac1n [ \sum_{i\in obs} (y_i  - \bar{y})^2 + \sum_{i\in lost} (y_i - \bar{y})^2 ]\nonumber\\
&=&\frac1n [ \sum_{i\in obs} (y_i  - \bar{y} \pm \bar{y}')^2 + \sum_{i\in lost} (y_i - \bar{y} \pm \bar{y}'')^2]\nonumber\\
&=&(1-l) {s'}^2 + l{s''}^ 2 + (1-l)(\bar{y}' - \bar{y})^2 + l(\bar{y}'' - \bar{y})^2\label{e:sbreakdown1}
\end{eqnarray}
where $l = \frac{n'}{n}$ is the sample loss rate.
By replacing $\bar{y}$ according to (\ref{e:meanbreakdown}), equation (\ref{e:sbreakdown1}) simplifies to:
\begin{equation}\label{e:sbreakdown2}
s^2 = (1-l) {s'}^2 + l{s''}^ 2 + l(1-l)(\bar{y}' - \bar{y}'')^2
\end{equation}
This breakdown holds for both treatment and control samples.
We will refer to them by $trt$ and $ctrl$ suffixes.

Each scenario must specify three unknown parameters: $\bar{y}_{ctrl}^{\prime\prime}$, $s_{ctrl}^{\prime\prime}$ and $\beta_{int}$.
The first two parameters are our hypothesis about the status of lost data points, regardless of experimentation.
For all simulations, we estimated these summary statistics from the bottom 10th percentile of quality metric distributions. 
This choice was motivated by our observation that most data loss is associated with poor experiences.
%
$\beta_{int}$ represents our assumption about the impact of the treatment on lost data.
This is established on a per-case basis for each experiment, using domain knowledge obtained from lab or offline results.
Large absolute values for $\beta_{int}$ are indicative of experiments with high sensitivity to data loss.
Setting $\beta_{int}$ to zero implies there is no difference in treatment effect on unobserved and observed samples, hence there is zero sensitivity to data loss.

Assuming that treatment may only change the metric baseline and not its variance, i.e. $s''_{ctrl} = s''_{trt}$ the three input parameters specified by a scenario are sufficient to reconstruct the overall treatment effect.
Using model (\ref{modelLoss}), $\bar{y}''$ can be estimated by:
\begin{equation}\label{e:unknown-mean}
\bar{y}''_{trt} = \bar{y}''_{ctrl} + \delta' + \beta_{int}
\end{equation} 
under scenario $[\bar{y}_{ctrl}'', s_{ctrl}'', \beta_{int}]$ where $\delta' = \bar{y}'_{trt} - \bar{y}'_{ctrl}$ is the observed delta.
The simulation process is a simple application of (\ref{e:meanbreakdown}), (\ref{e:sbreakdown2}) and (\ref{e:unknown-mean}), as described in Algorithm \ref{a:simulation1}.

If needed, the equal variance assumption can be simply relaxed by adding $s''_{trt}$ an extra input parameter to Algorithm \ref{a:simulation1} and applying it when calculating $se(\Delta)$. 
This parameter can be initiated using historical values from previous experiments. We continue with equal variance assumption since it is verified for our application. 

The result of applying Algorithm \ref{a:simulation1}  is a new scorecard with simulated values under no loss.
If there is a large difference between the simulated and observed scorecards, this implies that the data collected from the experiment are inconclusive.
On the other hand, the observed scorecard is deemed to be trustworthy enough if it is not practically different from the simulated scorecards.

Table~\ref{tab:expmobilevideo_sim} shows an example of the output of Algorithm \ref{a:simulation1}, for the \emph{Video Decoder} experiment under two different scenarios, along with observed statistics for a few metrics.
The experiment data used for this simulation include approximately 30 million observations for each variant.
As discussed in Section \ref{subs:loss_experimentation_examples}, the video telemetry in this experiment had over $10\%$ loss.
According to the observed scorecards, none of the metrics showed significant change.
The video-related metrics (shown in \emph{italic}) are calculated using high loss rate events, while the others are based on low loss rate events.
As expected, the latter set of metrics are the same in observed and simulated scorecards.
However, the video related metrics show wildly different results.
Under the worst-case scenario ($\beta_{int} = 5\%$), overall VideoFreezeDuration shows a $2.43\%$ increase, while under the best-case scenario ($\beta_{int} = -5\%$) it shows a $2.47\%$ decrease.
This indicates a large confidence interval for true effect, that spans over a diverse set of conclusions from ``treatment improves user experience'' to ``treatment degrades user experience''.
In such cases, we flag the experiment as inconclusive.

\begin{table}[t]
  \begin{center}\begin{tabular}{lccc}
    \multicolumn{4}{c}{Relative Delta (P.Value)}\\
    \hline
    \textbf{Metric} & \textbf{Observed} & \textbf{Best-case} & \textbf{Worst-case}\\
    \hline
    DurationMean & -0.02\% (0.76) & -0.02\% (0.76) & -0.02\% (0.76) \\
    CallEstablishRate & -0.04\% (0.30) & -0.04\% (0.30) & -0.04\% (0.32)  \\
    {\it VideoDuration} & 0.02\% (0.78) & 0.11\% (0.31) & 0.09\% (0.40) \\
    {\it VideoFreezeDuration} & -0.03\% (0.76) & -2.47\% (0.00) & 2.43\% (0.00) \\
    {\it VideoBitrateMean} & -0.04\% (0.78) & -0.54\% (0.00) & -0.68\% (0.00) \\
    \hline
    \end{tabular}
    \caption{Sample Experiment Scorecard - Video Decoder}
    \label{tab:expmobilevideo_sim}
    \end{center}
  \end{table}

  \setlength{\textfloatsep}{10pt}
  \begin{algorithm}[t]
  \caption{Detect Loss Tolerance }
  \begin{algorithmic}[1]
  \Require $\overrightarrow{l}$, $\overrightarrow{\delta''}$, significance level
  \Ensure low decision impact areas
  \ForAll {$(l, \delta'')$ pairs }
    \State Set input parameters of Algorithm \ref{a:simulation1} by $l_{trt} = l_{ctrl} = l$ and $\beta_{int} = \delta''$  
    \State Run Algorithm \ref{a:simulation1} to generate $\frac{\delta}{se(\delta)}$ 
    \State Generate p-value $p$ using standard normal distribution
  \EndFor
  
  \State Flag $\overrightarrow{l} \otimes \overrightarrow{\delta''}$ where p-values is greater than significance level as {\it safe} zone 
  \end{algorithmic}
  \label{a:simulation2}
  \end{algorithm}

\subsection{Data Loss Tolerance in Experimentation}
\label{subs:tolerance}
Applying Algorithm \ref{a:simulation1} to various \emph{Skype} experiments showed how much the actual treatment effect could be different from observed scorecards.
However, this difference only matters if it is large enough to impact our decision.
Often, decisions informed by AB tests are binary functions with two outcomes: ``roll-out'' or ``no roll-out''.
We call the data loss \emph{tolerable} if it does not reverse our decisions.
In this section, we discuss how to find the maximum tolerable data loss rate, which we refer to as the \emph{tolerance threshold}.
Note that tolerance threshold is irrelevant when correlation bias exists, since there is no tolerance for significantly different loss rates between treatment and control, as this leads to incomparable samples.
The solution provided in this section is used to find the data loss threshold where treatment and control have a statistically comparable level of loss (i.e., $l = l_{ctrl} = l_{trt}$).
 
The decisions made by experimentation are tightly bound to the p-values of outcome metrics provided by scorecards. 
A ``roll-out/no roll-out'' decision usually translates to the p-values of desired metrics being lower/higher than significance level\footnote{The significance level can be different for each organization depending the number of metrics they consider for experimentation and accepted level of false positive rate. A popular significance level for pairwise comparison on single metric is 0.05.}.
Leveraging the binary nature of experiment decision process, we approached this problem by comparing observed p-values with simulated p-values for complete samples (no data loss).
If both are on the same side of the significance level, data loss is tolerable for that experiment.
Without loss of generality, we set $\delta'=\bar{y}'_{trt} - \bar{y}'_{ctrl} = 0$.
This way, the observed p-value is higher than the significance threshold and we just need to track the simulated p-values.

To find a threshold that is tolerable for most experiments in a given experimentation platform, we need to run Algorithm \ref{a:simulation1} repeatedly for multiple loss rates $\overrightarrow{l}$, and under different scenarios $[\bar{y}_{ctrl}'', s_{ctrl}'', \beta_{int}]$.
The input parameters of Algorithm \ref{a:simulation1} must be adjusted to reflect the target population, the nature of data loss, and the sensitivity of experiments running on that platform. 
That is, to fix $[\bar{y}'_{ctrl}, s'_{ctrl}, s'_{trt}, \bar{y}'_{trt}]$ to the known summary statistics of the target population and to estimate $\bar{y}''_{ctrl}, s''_{ctrl}$ according to assumptions about the overall user experience under lossy conditions.

Therefore, most input parameters in Algorithm \ref{a:simulation1} are constant values reflecting the specifications of the experimentation platform, while $\beta_{int}$ varies over a possible range of values that represent different sensitivity levels for experiments run in the platform.
Since $\delta'$ is set to zero, $\beta_{int} = \bar{y}''_{trt} - \bar{y}''_{ctrl} = \delta''$ according to formula (\ref{e:unknown-mean}). 
We use $\delta''$ instead of $\beta_{int}$ for this simulation because it is easier to interpret for engineering teams, whose input is required for defining the range of $\beta_{int}$.

With the above settings, Algorithm \ref{a:simulation1} can be used to generate a 3-dimensional array of $[\overrightarrow{l}, \overrightarrow{\delta}'', \overrightarrow{p}]$ where $p$ is the simulated p-value for a complete sample based on given loss rate and experiment sensitivity level.


\setlength{\belowcaptionskip}{0pt}
\begin{figure}[t]
\centering
\includegraphics[width=3.5in]{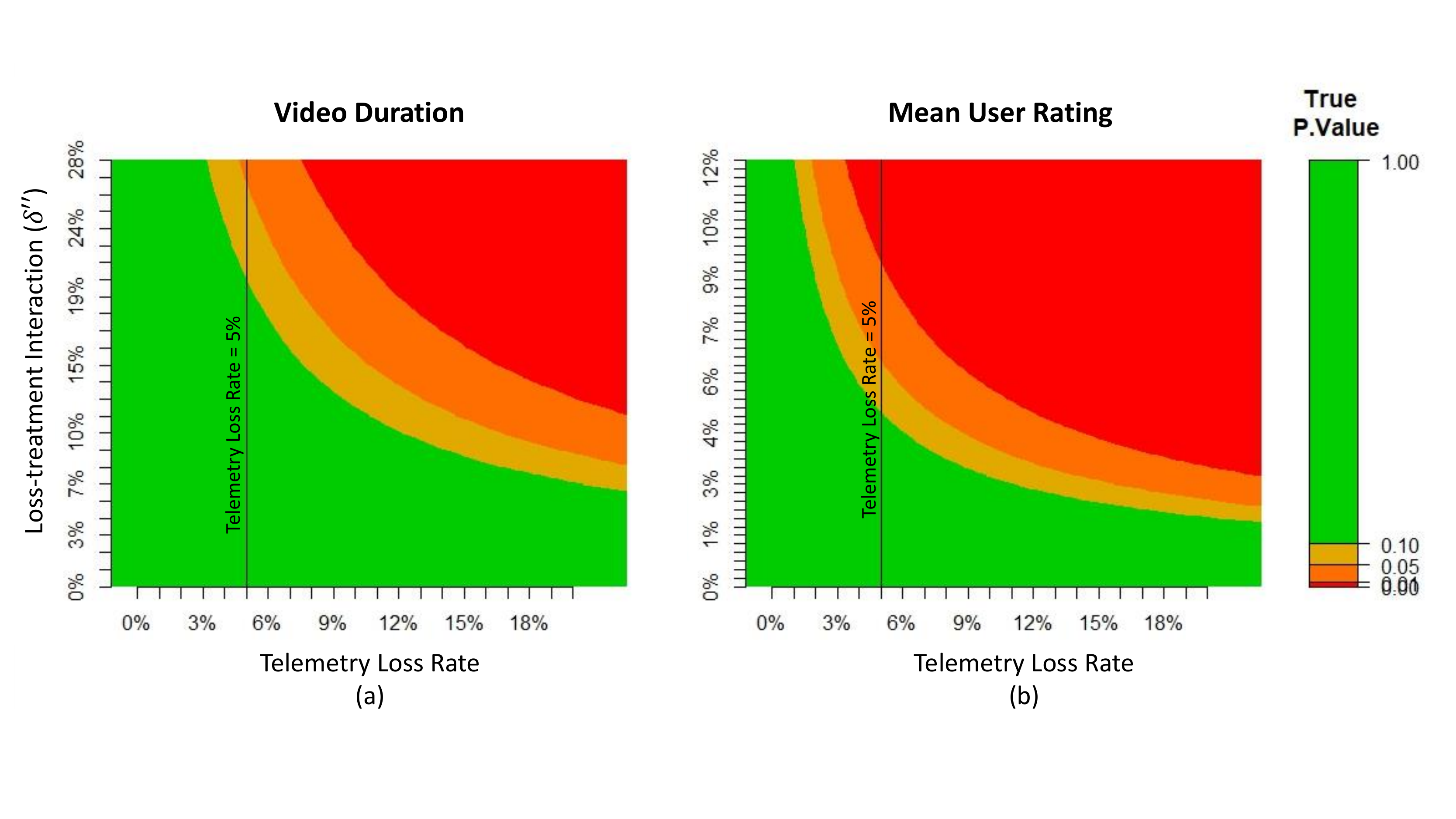}
\caption{Loss Tolerance plot for Video Duration on a low-end mobile platform (a); and Mean User Rating on a high-end mobile platform (b).}
\label{f:tolerance_plots}
\end{figure}

\noindent
\textbf{Discussion on setting input parameters:} Prior knowledge about the nature of data loss plays an important role in setting these parameters.
In \emph{Skype}, we found that the causes of data loss are related to poor experience.
A conservative choice is to set values drawn from the poor experience population (i.e., simulating $\bar{y}'_{ctrl}, s'_{ctrl}$ from the lower tail of outcome metric distribution).
Choosing values closer to the distribution center leads to a higher tolerance compatible with the MAR assumption.
It is also important to capture a realistic range for $\delta'' = \bar{y}''_{trt} - \bar{y}''_{ctrl}$ based on prior lab tests and metric noise levels. 
In the absence of such knowledge, $(0, 2\times S)$ where $S$ is metric standard deviation, is a reasonable range.

\noindent
\textbf{Application to \emph{Skype} Experimentation:}
At \emph{Skype}, we applied Algorithm \ref{a:simulation2} to various segments of the target population defined by device platforms (e.g., Android, Windows) due to the heterogeneity of metric noise levels and baselines across different platforms.
%
\noindent
The following provides further detail on the input parameters:
\begin{enumerate}[leftmargin=*,noitemsep,topsep=1pt]
   \item The summary statistics of lost data points in control group are set to the lower 10th percentile of the respective metrics
   \item $\overrightarrow{l}$ ranges from 0 to $20\%$ to cover all possible loss rates
   \item $ \overrightarrow{\delta''}$ ranges from 0 to the minimum of $30\%$ of metric average and $50\%$ of its standard deviation
   \item The significance level is set to 0.01
\end{enumerate}

As shown in Figure \ref{f:tolerance_plots}, higher loss rates increase the chance of decisions being reversed.
However, the rate of increments varies by metric and platform.
Average User Rating on a high-end mobile platform, for example, has higher sensitivity to data loss compared to Video Duration on a mobile low-end platform.
This is shown as larger safe zone (green area) for data loss on Video Duration.

These two examples show how variable the tolerance could be depending on experiments' sensitivity to loss.
Experiments that are designed to enhance user experience in poor network conditions, for example, may fall in the top area in these plots and have lower tolerance for data loss.
In practice, there is no unique non-zero tolerance threshold that uniformly satisfies all types of experiments.
We set the overall target for data loss reduction to the threshold value that would ensure trustworthy experiments for majority of treatment types. 
According to the simulation results, $5\%$ is the threshold value that provides tolerable coverage and risk of false positives/negatives for \emph{Skype}.
 
%
%
%

\section{Measuring Data Loss}
In this section, we first present two approaches to measure loss of telemetry events.
Then, we evaluate these techniques in practice, using the results from various experiments.
Finally, we provide some recommendations on how telemetry loss can be reduced.

\label{s:dataloss}

\subsection{Anchor Method}
\label{subs:anchor_method}
This method relies on pairing (or anchoring) client events with a more reliable server event. 
Call establishment in \emph{Skype} is negotiated using a server.
Therefore, for every call (uniquely identified by $call\_id$), a record of the call's technical telemetry and the participating legs is reported by the server.
Server machines are reliable, not bandwidth-constrained, and therefore exhibit near-zero telemetry loss.
The unreliable client events can therefore be paired with the highly reliable server events to get an estimate of the loss in client telemetry.
Each leg of the call is uniquely identified by $leg\_id = (call\_id, endpoint\_id)$

\begin{figure}[t]
\begin{center}
  \includegraphics[width=\linewidth]{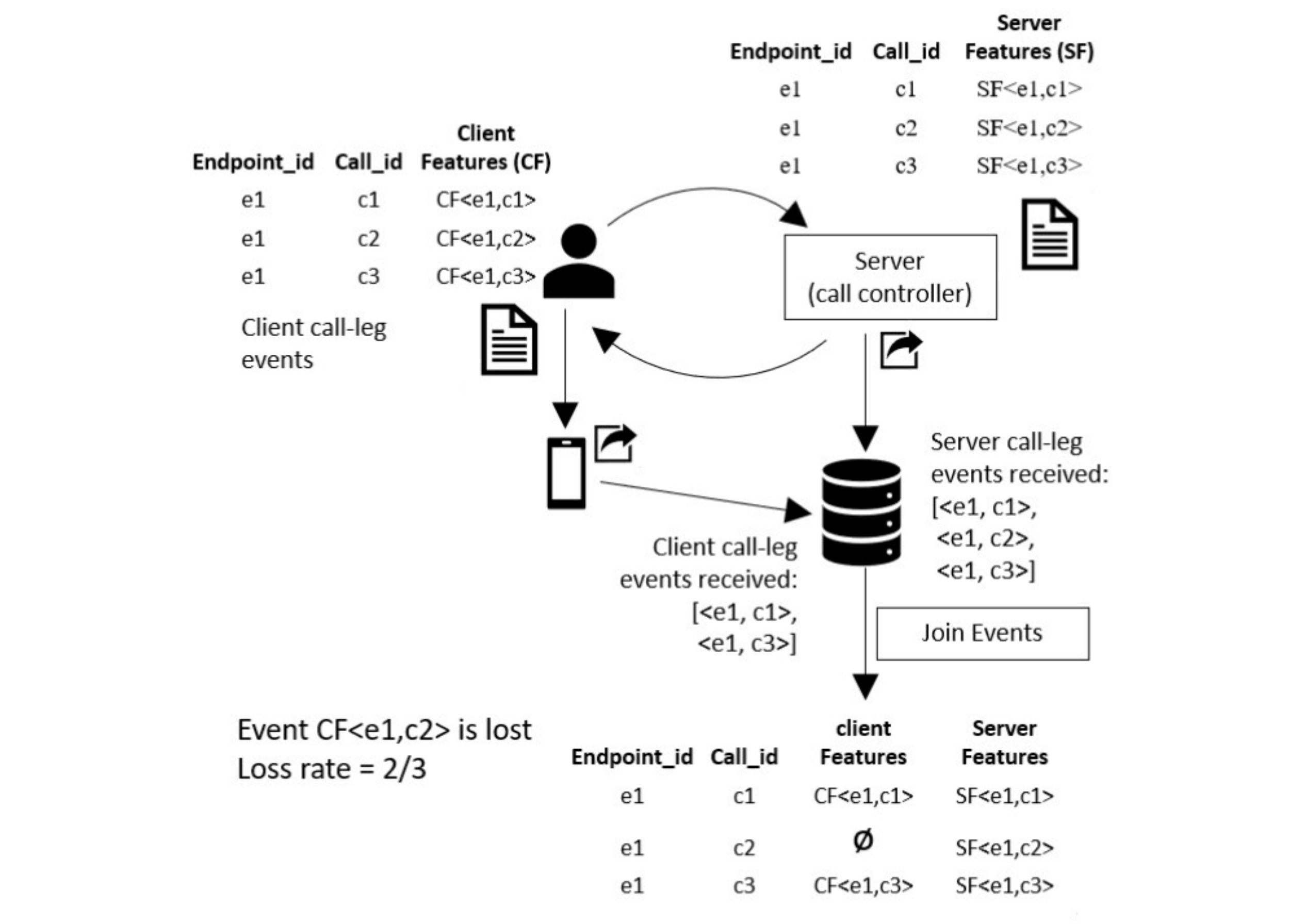}
  \caption{The Anchor method to estimate data loss.}
  \label{f:anchor-method}
\end{center}
\end{figure}

The anchor method is shown in Figure \ref{f:anchor-method} and Algorithm \ref{a:anchor-method}. 
Figure \ref{f:anchor-method} shows the table of events reported by the client and server, where each row corresponds to a call. 
Note that the figure only shows one leg of the call.
For each of these call legs, the server also logs the leg IDs, allowing the client and server leg events to be paired.
In this example, measures of event $CF<e1, c2>$ 
 could not be uploaded and this constitutes a loss.

Note that the anchor method is limited to scenarios where both client and server events are submitted.
For example, loss can be measured only for call attempts where the request/acknowledgment reached the server.
Therefore, this method can be applied to all \emph{established} calls but not all \emph{attempted} calls. 

\setlength{\belowcaptionskip}{-1pt}

\setlength{\textfloatsep}{10pt}
\begin{algorithm}[t]
\begin{algorithmic}[1]\scriptsize
\Statex
\Procedure{AnchorMethod}{$server\_events\_set, client\_events\_set$}
  \State $events\_lost = 0$
  \State $expected\_events = \Call{Count}{server\_events\_set}$

  \Statex
  \ForAll {$leg\_id \in server\_events\_set$}
		\If{$\Call{Not}{ \Call{HasKey}{client\_events\_set, leg\_id}} $}
			\State $events\_lost \gets events\_lost + 1$
		\EndIf
  \EndFor	

  \Statex
  \State $anchor\_loss\_rate  \gets \frac{events\_lost}{expected\_events}$
  \State \textbf{return $anchor\_loss\_rate$} 
\EndProcedure


\end{algorithmic}
\caption{Data loss estimation using the anchor method.}

\label{a:anchor-method}
\end{algorithm}

\subsection{Sequence Method}
\label{subs:sequence_method}
\begin{figure}[t]
\begin{center}
  \includegraphics[width=0.9\linewidth]{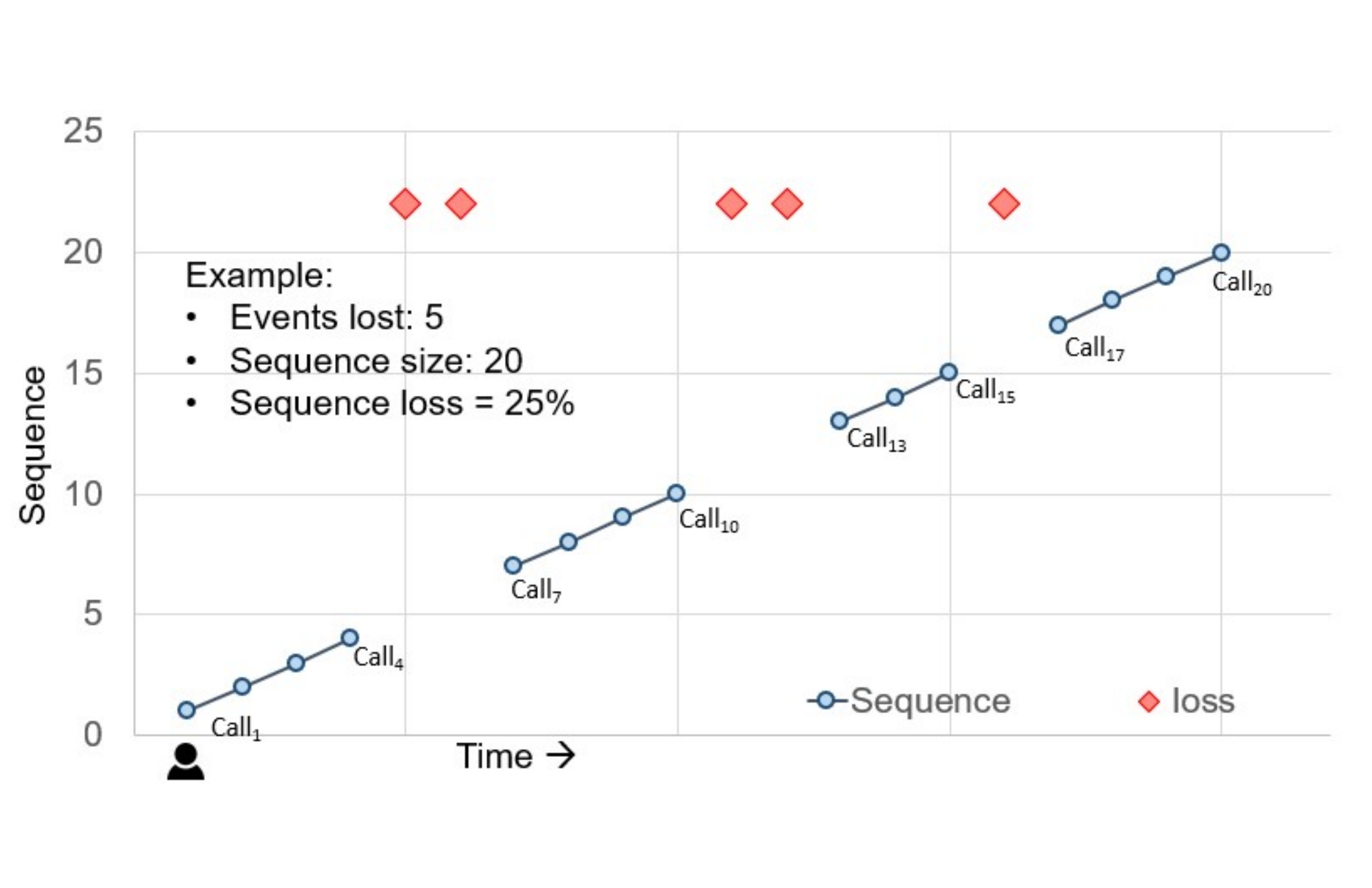}
  \caption{The Sequence method to estimate data loss.}
  \label{f:sequence-method}
\end{center}
\end{figure}

The sequence method (Algorithm \ref{a:sequence-method}) is a more general solution compared to the anchor method, 
using a monotonically increasing counter that is persisted to the client's local storage.
This counter is referred to as a \emph{sequence number} ($sn$).
Each event has an associated counter per endpoint to track the number of events that have been generated.
This monotonically increasing set is referred to as a \emph{sequence}, as shown in Figure \ref{f:sequence-method}.
At the beginning of each call, the $sn$ value for each event is incremented and reported as part of the event.
After the telemetry uploader transmits the client events, the associated $sn$ values can be reconstructed.
If there is no loss in events, there should be no gaps in the $sn$ values when considered in sorted order; a gap in $sn$ indicates that a client event has been lost, with the size of the gap indicating the number of lost events.

Algorithm \ref{a:sequence-method} shows how to compute the sequence loss for a batch of de-duplicated events.
In practice, in a live system, the backend system needs to maintain a lookup table of the sequence information.
Specifically, the lookup table needs to maintain the following counters:
1) last received sequence number, $prev\_sn$;
2) the cumulative loss so far, $sequence\_gap$; and 
3) the size of the sequence, $expected\_sequence\_size$.
By accumulating the $sequence\_gap$ values and $expected\_sequence\_size$ values across all sequences, we can incrementally compute the overall data loss.

The sequence method can be applied in situations where there is no reliable server method to pair with, such as randomly sampled events that do not have a reliable anchor (e.g., user ratings).
As a practical matter, $sn$ values are reset when users uninstall and reinstall the app (common in low-end devices as users try to conserve space by removing applications when not needed).
While this is a corner case, these resets need to be detected and handled properly.

\setlength{\belowcaptionskip}{-5pt}

\setlength{\textfloatsep}{6pt}
\begin{algorithm}[t]
\begin{algorithmic}[1]\scriptsize
\Require  $MIN\_SEQUENCE\_SIZE$
\Statex

\Procedure{SequenceMethod}{$sequence\_list$}
\State $events\_lost \gets 0$
\State $events \gets 0$

\ForAll {$sequence \in sequence\_list$}
	\State $(seq\_gap, seq\_size) \gets \Call{SequenceLoss}{sequence}$
	\State $events\_lost \gets events\_lost + seq\_gap$
	\State $expected\_events \gets expected\_events + seq\_size$
\EndFor

\Statex

  \State $sequence\_loss\_rate  \gets \frac{events\_lost}{expected\_events}$
  \State \textbf{return $sequence\_loss\_rate$} 
\EndProcedure

\Statex
\Statex

\Procedure{SequenceLoss}{$sequence$}
 \State $expected\_sequence\_size \gets \Call{Max}{sequence} - \Call{Min}{sequence} + 1$
  \If{$expected\_sequence\_size < MIN\_SEQUENCE\_SIZE$}
	\State \textbf{return $(0, 0)$}
  \EndIf

  \Statex

  \State $sequence\_gap \gets expected\_sequence\_size - \Call{Count}{sequence}$





  \State \textbf{return $(sequence\_gap, expected\_sequence\_size)$} 
\EndProcedure

\end{algorithmic}
\caption{Data loss estimation using the sequence method.}
\label{a:sequence-method}
\end{algorithm}

\subsection{Data Loss Measurements in Practice}
\label{subs:data_loss_results}

We use the \emph{CST} event described in Section \ref{subs:telemetry_loss} to show the results of the anchor and sequence methods and discuss their practical implications.
Although we cannot report absolute numbers due to confidentiality restrictions, we show appropriately scaled relative values to convey the results.
Since the anchor method uses server-side events as the baseline, we compare the two methods only for calls that are established (i.e., the server has a record of the caller's attempt or callee's response).
We set the $MIN\_SEQUENCE\_SIZE$ to $5$, which we will explain in further detail later.
The dataset used for the analysis consisted of more than a billion established \emph{Skype} calls over a period of several weeks.
The anchor and sequence methods were implemented using Microsoft's big data analysis platform \cite{chaiken2008scope}, which has a query language is similar to Apache Hive \cite{thusoo2009hive}.
Our implementation is part of the production telemetry and experimentation processing pipeline.

\begin{figure}[t]
  \centering
  \begin{subfigure}[b]{0.49\linewidth}
    \centering
    \includegraphics[width=\textwidth]{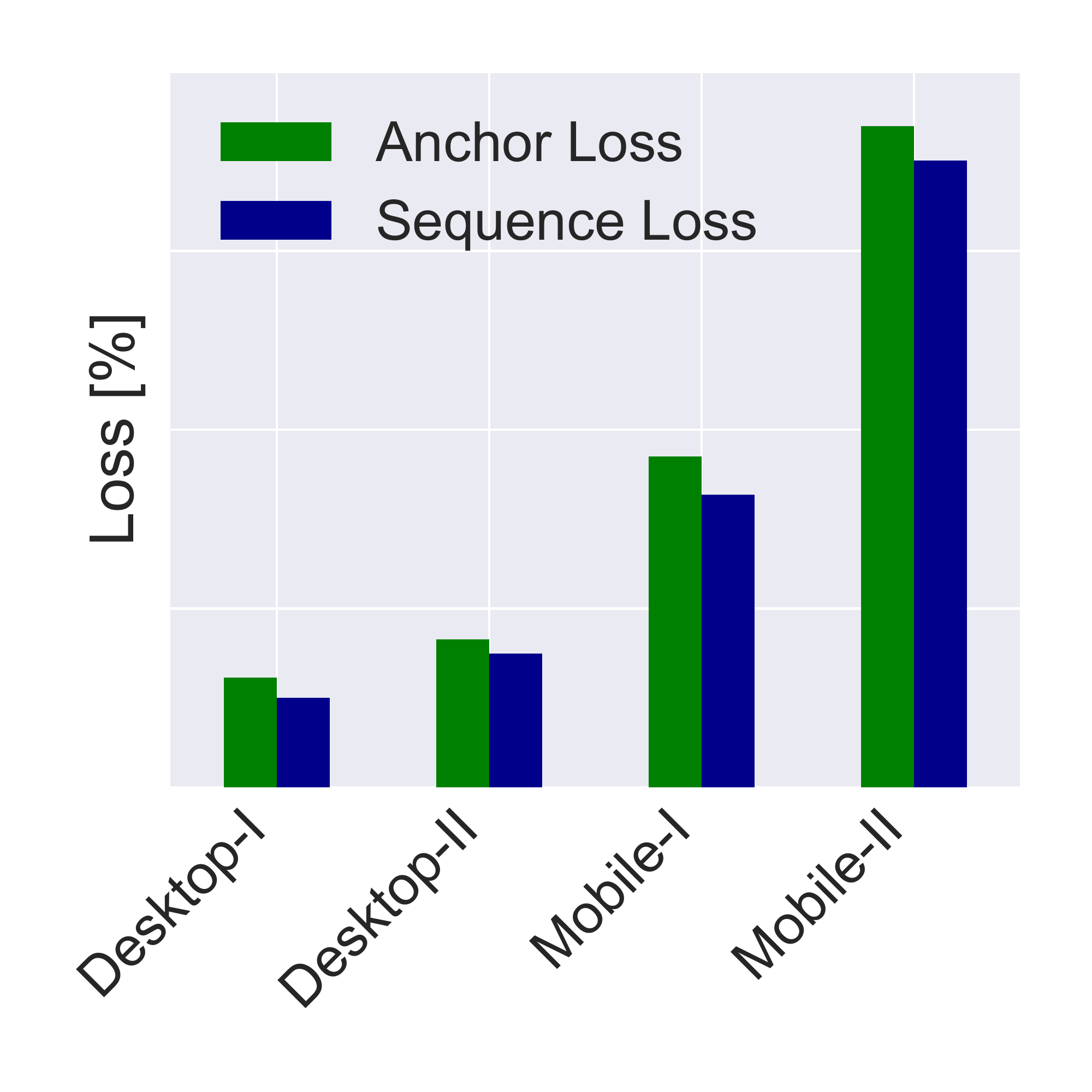}
    \caption{} 
    \label{f:compare_data_loss}
  \end{subfigure}
  \begin{subfigure}[b]{0.4675\linewidth}
    \includegraphics[width=\textwidth]{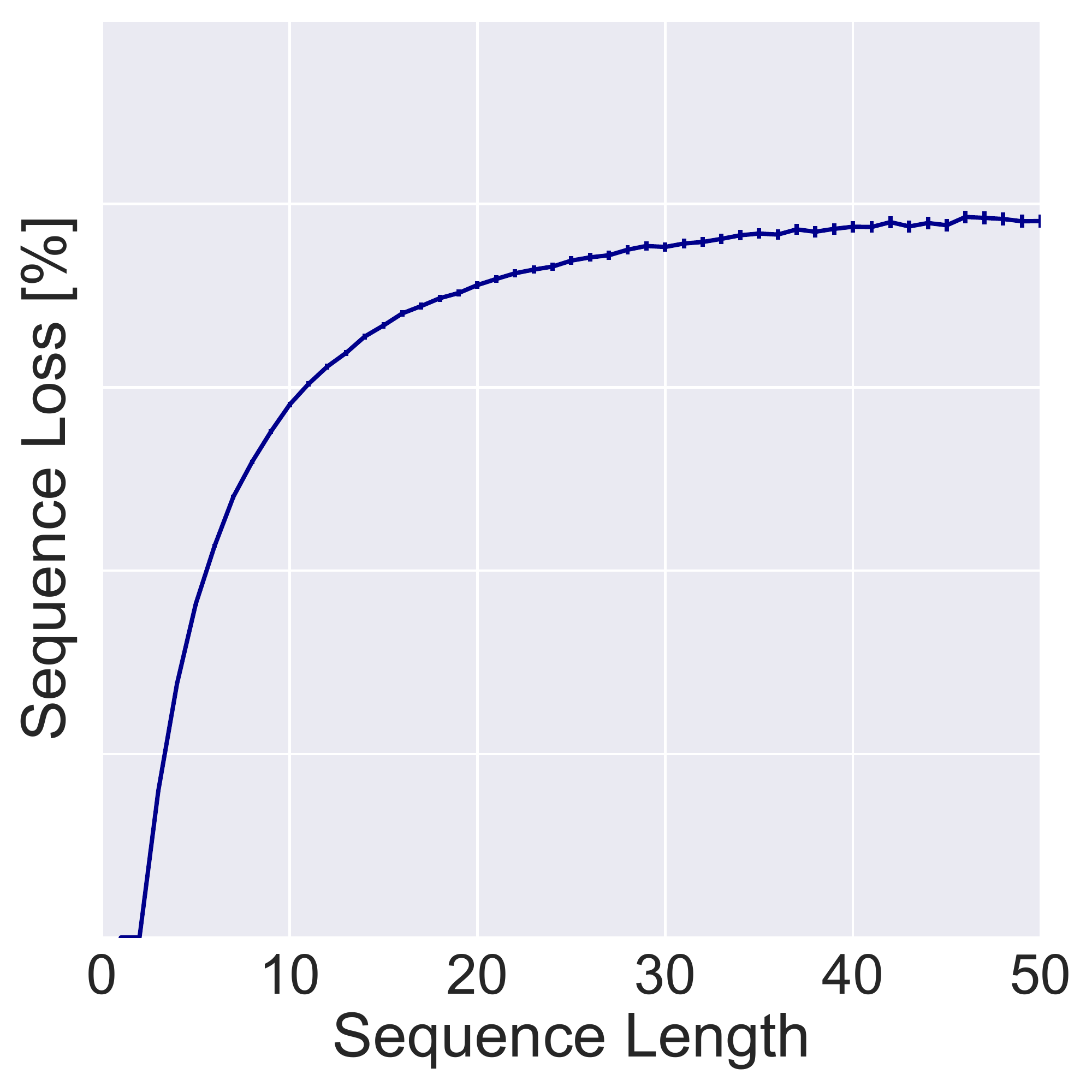}
    \caption{}
    \label{f:sequence_loss_size}
  \end{subfigure}
  \caption{
  (a) Estimates of data loss in the \emph{CST} event using the 
  Anchor and Sequence methods. 
  (b) Impact of sequence length on sequence loss estimates for mobile platforms.}
  \label{fig:test}
\end{figure}

The loss rate of the \emph{CST} client side event as estimated by the anchor and sequence methods is shown in Figure \ref{f:compare_data_loss}.
The overall absolute difference between the two methods is less than $0.5\%$.
Not surprisingly, the absolute loss is lower for desktop platforms than mobile platforms.
From our dataset, we found that the estimated difference in loss between the two methods is less than $0.3\%$ for desktop platforms, and about $0.7\%$ for mobile platforms.
Due to the large number of events, the confidence intervals of the estimates are very small and not shown in the figure.
Note that that the Sequence method consistently estimates lower loss than the Anchor method.

\setlength{\belowcaptionskip}{5pt}

The length of each sequence is a function of the number of calls made by the endpoint during the time period used for the study.
In general, the distribution of the sequence size will vary per application, based on the usage characteristics.
Since we have millions of users making calls at varying levels of activity, we can study the relationship between sequence size and sequence loss.
Figure \ref{f:sequence_loss_size} shows this relationship for our mobile platforms, showing that the estimates converge as the size of the sequence increases.
This is because the uncertainty in the loss estimate of the sequence endpoints decreases as the sequence grows.
Due to this effect, the sequence method tends to underestimate the overall level of loss when sequences are short.
In our experience, computing sequence loss for sequences of size greater than 5 provides a good trade-off between coverage and accuracy.


\subsection{Discussion on Data Loss Estimation}
\label{subs:data_loss_estimation}
The practical tradeoffs and qualitative comparisons between the sequence method and anchor method are shown in Table \ref{tab:dloss_pros_cons}.

\begin{table}
  \caption{Comparison of Anchor and Sequence methods} 
\begin{center}
  \begin{tabular}{|p{0.19\linewidth} | p{0.33\linewidth} | p{0.35\linewidth}|}
  \hline
  \textbf{Topic} & \textbf{Anchor Method} & \textbf{Sequence Method} \\ [0.5ex] 
  \hline
  Dependencies &Server side event required as a source of ground truth. &Relies purely on client side telemetry.  \\ 
  \hline
  Sampled events &Typically, a good anchor is not available. &Can be used for all events. \\
  \hline
  Accuracy &Reliable ground truth results in high accuracy. &More accurate for longer sequences. \\
  \hline
  Treatment effect bias &Loss estimates do not lead to biases between control and treatment. &Loss estimates may lead to biased comparison due to $ {\scriptstyle MIN\_SEQUENCE\_SIZE } $ threshold. \\  
  \hline
  State maintenance &The approach is stateless.	&Table of sequence information needs to be maintained for incremental processing. \\
  \hline
  Integration with experimentation scorecard &Easier to integrate with the scorecard due to its simplicity. &Integration with scorecard is 
  more challenging due to the state overhead. \\
  \hline
\end{tabular}
 \label{tab:dloss_pros_cons}
 \end{center}
\end{table}

For the majority of the scenarios in VoIP calling, the anchor method is appropriate since we have a server side event to anchor with.
Moreover, the anchor method provides better accuracy and requires less state maintenance.
As a consequence, the anchor method is easier to integrate in experimentation scorecards.
However, the timescale of many of the experiments conducted in \emph{Skype} last in the order of days.
At this timescale, the sequence method may not provide adequate coverage, due to a large number of sequences falling below the $MIN\_SEQUENCE\_SIZE$ threshold.
In our experiments, at least two weeks worth of data is required for sequence method to provide adequate coverage.
Nevertheless, the sequence method has been used to measure loss rate of events where a good anchor is not available.
Even if sequence method cannot be easily integrated in the experimentation scorecard, it can be used to establish the overall loss rate of events, and in-turn, improve the trustworthiness of the metrics derived from those events.

\subsection{Best Practices for Reducing Data Loss}
\label{subs:loss_best_practices}
In our experiences with \emph{Skype}, we have successfully improved event reliability without impacting service quality using a system that is service aware, with prioritized, persistent event queues.
This has helped us bring the loss to below $2\%$ for events used in our experimentation system.
While the design of the telemetry system is beyond the scope of this paper, we would like to share some of the lessons we have learned on low-loss telemetry design, including how careful design of telemetry events can make the best use of the available client resources (bandwidth, storage, etc.):
%
%
\begin{itemize}[leftmargin=*,noitemsep,topsep=1pt]
\setlength\itemsep{0.2em} 
\item{KPI hierarchy design:
Organizations need to carefully design a system of metrics with a clear hierarchy. We recommend three tiers. 
Tier-0 represent business health metrics, Tier-1 represents leading indicators of quality/reliability, and Tier-2 represents operational metrics of sub-components.
The events should be designed and prioritized using this map.
This ensures that the most informative events suffer the least loss.
In an A/B test aimed at evaluating the impact of prioritization for one the business-critical KPIs, we found an absolute loss reduction of $3.8\%$ when the priority of the event was increased by one level.
}
\item{Split large events:
Keeping the size of each event small is critical to minimize congestion.
Moreover, the measures in each event should provide information for metrics at the same tier.
For example, mixing high priority measures/metrics (e.g., user ratings) with lower priority ones (e.g., UI selections) will cause unnecessary loss in critical information.
In a lab experiment, we found that lowering the event size from 24 KB to 3 KB resulted in reducing the loss from $30\%$ to $4\%$.
}
\item{Review feature importance:
Over the life-cycle of a product, architectural changes and bug fixes lead to changes in the usefulness of measures.
It is critical to periodically evaluate the importance of these measures by correlating them with a Tier-0 or Tier-1 metric. Measures with limited correlation should be either removed from the event or investigated to keep the size small.
}
\end{itemize}

\noindent
While we cannot completely eliminate data loss, measuring its extent and applying the above principles to minimize it are crucial prerequisites for trustworthy experimentation.

\section{Practical Guidance for Experimenters}
\label{s:guidance}
Our experiences of running experiments have taught us that the impact of data loss can be very significant,
and its importance can be easily underestimated.
Therefore, we want to share some general guidance for the benefit of the online experimentation community.
We have followed an iterative process in improving the trustworthiness of our experimentation system and its resilience to telemetry loss, as described in the steps below:
\begin{enumerate}[leftmargin=*,noitemsep,topsep=1pt]
  \setlength\itemsep{0.2em} 
  \item Measure and track data loss for relevant events by adding them to the experimentation scorecard using the methods described in Section \ref{s:dataloss}.
  \item Estimate how much loss that can be tolerated by the experimentation system using the methodology described in Section \ref{s:boundaries}, and exclude telemetry events that suffer higher levels of loss from decision making.
  \item Use examples similar to Section \ref{subs:loss_experimentation_examples} to communicate biases introduced by these losses.
  \item Mitigate the impact of loss by restructuring events using recommendations outlined in Section \ref{subs:loss_best_practices}.
  \item Investigate the source of losses for events, starting with those with the highest loss rates.
\end{enumerate}

In our experience, we found that it is critical to make the estimates of loss rates part of the experiment scorecard, reporting these alongside the other metrics monitored by the experimenter.
By doing this, it becomes easier for experimenters to detect biases, and provides an extra check for the validity of experimental results.
In our organization, we flag events that suffer a loss rates greater than $5\%$, and mark metrics relying on these events as invalid.
For example, for a given experiment, suppose the loss in \emph{CST} event was higher than $5\%$; in this case, the $call\_duration$ metric will be marked 
invalid for the purposes of analysis and conclusions.

We also find this $5\%$ threshold useful when considering newly added metrics, since new events often exhibit high loss rates ($10\%$ or more).
In such cases, we ask the appropriate engineering team to find the root cause and fix bugs as needed (e.g., the \emph{ui-change} experiment 
discussed in Section \ref{s:introduction}), to bring the loss rate below the allowed threshold before they can be used in any experimentation analysis.
Several teams have gone through this exercise.
Their focused efforts have helped lower the levels of telemetry loss, and improved the overall trustworthiness of our online experiments.

\section{Conclusion}
\label{s:conclude}
Telemetry loss is an inseparable part of online experimentation with potentially dramatic implications.
Since many apps are required to operate under resource constraints and challenging network environments, this problem is
even more severe in client experiments.
Based on several examples observed from real experiments at scale, we show the impact of telemetry loss on experiment outcomes.
In this paper, we argue that it is vital to measure and track the level of loss in experimentation systems.
We provide a theoretical framework for characterizing the types of biases introduced by telemetry loss.
Using this framework, we provide a methodology for experimenters to evaluate the amount of loss that can be tolerated in their systems.
To measure the absolute level of loss, we present the anchor method and sequence method: two practical approaches that have been deployed at scale.
While we note that completely eliminating telemetry loss is not practical, we present the community with a set of best practices to reduce loss rates and manage the problem.
These methods have already been adopted by several applications with millions of users across billions of sessions.
Finally, we would like to emphasize that these principles can be applied generally, to improve the trustworthiness of any online 
experimentation system running at scale.

\section*{Acknowledgments}
\label{s:acknowledgments}
We thank Sergey Sukhanov for the sequence number implementation and Sooraj Kuttykrishnan for his useful comments and feedback.
We also thank the data pipeline, experimentation, and telemetry management teams within Skype/Microsoft; without their significant contributions, we would be flying blind to telemetry loss.

\end{document}